\documentclass[5p]{elsarticle}
\usepackage{amsmath,amssymb}
\newcounter{bla}

\begin{document}
\sloppy
\begin{frontmatter}

\title{Probability of graphs with large spectral gap by multicanonical Monte Carlo}
\author[a]{Nen Saito}
\author[b]{Yukito Iba}
\address[a]{Graduate School of Science and Cybermedia Center, Osaka University,
Toyonaka, Osaka 560-0043, Japan}
\address[b]{The Institute of Statistical Mathematics,
10-3 Midorimachi, Tachikawa, Tokyo 190-8562, Japan}

\begin{abstract}
Graphs with large spectral gap are important in various fields such as
biology, sociology and computer science.
In designing such graphs, an important question is how the probability of
graphs with large spectral gap behaves.
A method based on multicanonical Monte Carlo
is introduced to quantify the behavior of this probability,
which enables us to calculate extreme tails of the
distribution.
The proposed method is successfully applied to random $3$-regular graphs
and large deviation probability is estimated.

\end{abstract}

\end{frontmatter}

\section{Introduction}
Random graphs often appear and have been studied in various fields
of natural science and engineering.
A recent interest in their applications is generating expander graphs~\cite{sarnak2004expander},
regular graphs that shows high connectivity and homogeneity.
Such graphs have important applications
in designing networks of computers~\cite{gkantsidis2006random}, infrastructures~\cite{barthe'lemy2006optimal}, and real and artificial 
neurons~\cite{kim2004performance, donetti2005entangled}.

A way to define and generate expanders
is maximization of the spectral gap $g$, which is defined
as the difference between the
largest eigenvalue and the second largest eigenvalue of the adjacency matrix
of a graph. Donetti {\it et al.}~\cite{donetti2005entangled,donetti2006optimal}
numerically maximized the spectral gap by simulated
annealing and generated examples of these graphs.
However, in designing networks, we are also interested in
quantitative properties; specifically 
how the probability of large spectral gap graphs behaves when the
size $N$ of graphs increases.

In this paper we apply a method based on multicanonical Monte Carlo~\cite{berg1991multicanonical, berg1992new} to
the calculation of large deviations in the spectral gap of random
graphs. The method can be regarded as an extension
of the method introduced in~\cite{saito2010multicanonical_arXiv}; in~\cite{saito2010multicanonical_arXiv}, 
large deviations in the largest eigenvalue of random matrices 
are computed by a similar method.

Multicanonical Monte Carlo enables us to estimate tails of the
distribution whose probability is very small and cannot be
computed by naive random sampling.
By using the proposed method, we estimate the distribution of the spectral gap of
random 3-regular graphs and quantify the probability 
$P(g >\xi)$ that the spectral gap is larger than a given $\xi$;
when $\xi \gtrsim 0.18$, it is
shown that $P(g>\xi) \sim \exp(-N^2 \Phi (\xi))$ for large $N$.
\section{Spectral Gaps}
An undirected graph is described by the corresponding adjacency matrix $A_{ij}$, whose entries
are defined by
\begin{equation}
 A_{ij}= \left\{
\begin{array}{cc}
1 & \mbox{$i$ and $j$ are connected,} \\
0 & \mbox{otherwise.}\\
\end{array}
\right.
\end{equation}
Here we denote eigenvalues of the adjacency matrix by $\{ \lambda _i \}$.
In the case of $k$-regular graphs, $A_{ij}$ has the trivial largest eigenvalue
$\lambda _1 =k$, therefore we assume 
\begin{equation}
k=\lambda _1 \ge \lambda _2 \ge ... \ge \lambda _N.
\end{equation}
The difference $\lambda _1 - \lambda _2$ between the
largest eigenvalue and the largest non-trivial eigenvalue 
 of $A_{ij}$ is called ``spectral gap'' and takes a
non-zero value if the corresponding graph is connected.
We are interested in tails of the distribution of the spectral gap.
In the case of regular graphs,
Alon and Boppana~(see~\cite{alon1986eigenvalues}) proved an asymptotic lower bound of the largest non-trivial
eigenvalue $\lambda _2$ as
\begin{equation}
 \lim_{\,\,\,\,\,\,\,N \to \infty} \!\!\!\!\!\!  \inf \lambda _2 \ge 2\sqrt{k-1}.
\end{equation}
Assuming that the limit exists, this gives an asymptotic upper bound for the spectral gap as 
$ \lim_{N \to \infty} g \le k-2\sqrt{k-1}$.
Graphs with $g \ge k-2\sqrt{k-1}$ are called ``Ramanujan
graphs''~\cite{lubotzky1988ramanujan}.
On the other hand, Friedman~\cite{friedman2003proof} proved that for
$k\ge 3$ and for any constant $\epsilon>0$ ``most''
random $k$-regular graphs have $\lambda _2$ that satisfies
$\lambda _2 \le 2\sqrt{k-1}+\epsilon$ as
$N \to \infty$. These results indicate that the peak of the
distribution of $\lambda_2$ is located near $2\sqrt{k-1}$ for large $N$.
Miller {\it et al.}~\cite{miller2008distribution} studied the
distribution around this peak using naive random sampling.
None of these studies, however, discusses extreme tails of the distribution of $\lambda_{2}$
or $g$, which is the main subject of this paper.
\section{Methods}
Here, we will give a brief explanation of multicanonical Monte Carlo.
The aim of this method is to estimate the density of states
$\Omega (g)$ defined by
\begin{equation}
 \Omega(g)=\int \delta (g(A)-g)\, dA,
\end{equation}
where $\delta$ is the Dirac $\delta$-function and $\int \mbox{d}A$
denotes a multiple integral in the space of matrix $A$.
We define the weight by $w(g)$ as a function of $g$, a key quantity of
this method.
When a weight function $w(g)$ is given, we
can generate samples from the distribution defined
by the weight using the Metropolis
algorithm~\cite{metropolis1953equation, landau2005guide}.
The essential idea is to tune the weight function $w(g)$ so as to produce a flat histogram of $g$.
After we find a appropriate weight function $w^{*}(g)$ that gives a flat
histogram, an approximate value of the density of states is estimated by $1/w^{*}(g)$.
To obtain this $w^{*}(g)$, we modify the weight function step-by-step through 
the Metropolis simulation
using the current guess of the weight function $w(g)$.
There are several ways to modify the function $w(g)$.
Among them, a method proposed 
by Wang and Landau~\cite{wang2001efficient}
is most successful
and used in this study.
For a more accurate estimate of the density of states, we calculate a histogram
$h^{*}(g)$ by performing a long simulation with fixed $w^{*}(g)$.
Then $\Omega (g)$ is estimated by $\Omega (g) \propto h^{*}(g)/w^{*}(g)$
and the probability distribution $p(g)$ of $g$ is obtained by $p(g)=\Omega(g)/\sum_{g}{\Omega(g)}$.
Practically, we calculate the
density $p(g)$ only in a prescribed interval 
$g ^{min} < g < g ^{max}$.

Details of the implementation of the Metropolis
algorithm are as follows.
The simulation starts from an arbitrary
$k$-regular graph with the desired number of
vertices $N$.
In each step, a candidate $A^{new}$ is generated 
by rewiring edges in the way used
in~\cite{kim2004performance,donetti2006optimal,maslov2002specificity},
where the degree $k$ of each vertex 
is not changed; a pair of
links $ij$ and $kl$ that satisfy $A_{ij}=A_{kl}=1, A_{ik}=A_{jl}=0$ is selected and
rewired as $A_{ij}=A_{kl}=0, A_{ik}=A_{jl}=1$.
Then the spectral gap $g$ of a candidate is calculated by
the Householder method and accept/reject decision of the
transition from the current state $A^{old}$ to
$A^{new}$ is made by comparing the Metropolis ratio
\begin{equation}
 \alpha = \frac{w(g(A^{new}))}{w(g(A^{old}))}.
\end{equation}
with a random number uniformly distributed in $(0,1]$.
A candidate with $g =0$, indicating
a disconnected graph, is always rejected,
hence an ensemble of connected random $k$-regular graphs is sampled.
\section{Results}
\begin{figure}[t]
\begin{center}
\includegraphics[width=0.40\textwidth]{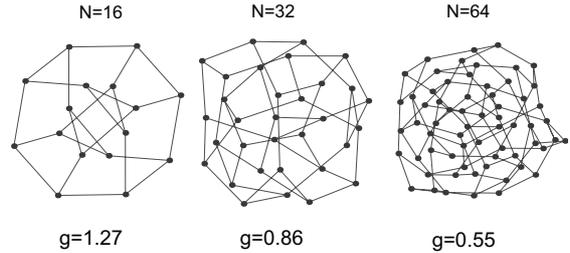}
\small
  \caption{3-regular graphs with the largest spectral gap found in the
 simulation.}
  \label{fig:1}
\end{center}  
\end{figure}

\begin{figure}
\begin{center}
\includegraphics{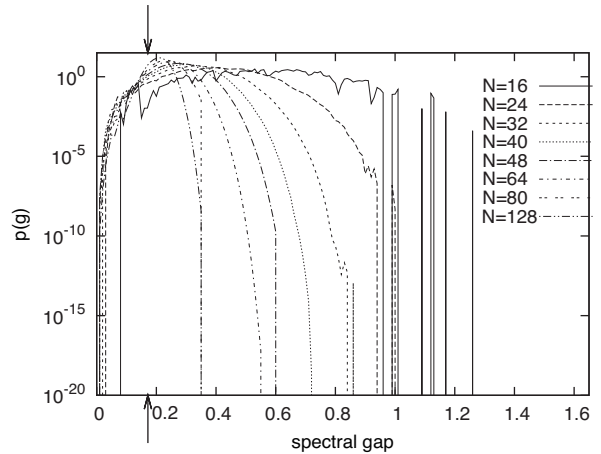}
\small
  \caption{The estimated probability density $p(g)$ of the spectral gap of
random 3-regular graphs. The density at $g$ plotted in the vertical
axis is obtained from the probability in a small bin around $g$ of a fixed width;
the values of them depend on the bin width in the right tail of the distribution, 
because discreteness of the spectra becomes relevant there. The arrows
indicate $g=3-2\sqrt{3-1}$;
graphs whose $g$ is above this value are Ramanujan.
}
  \label{fig:2}
\end{center}  
\end{figure}
Using the proposed method, we estimate the distribution of the spectral gap of random 3-regular graphs with the
number of vertices $N \le 128$.
Figure~\ref{fig:1} shows graphs with the largest spectral gap for
$N=16,32$ and $64$ found in the simulations.
In Figure~\ref{fig:2}, we show $p(g)$ of random 3-regular graphs with the
number of vertices $16 \le N \le 128$;
the computational time is 7 hours for $N=16$ and 251 hours for $N=128$ using a core of Intel Xeon X5365.
As $N$ increases, the probability density becomes sharper and the peak becomes
closer to around $3-2\sqrt{3-1}=0.172$, which is
consistent with a theoretical estimate~\cite{friedman2003proof} and
a numerical experiment~\cite{miller2008distribution}.
Specifically, the probability of graphs with large $g$ decreases
drastically for large $N$.

\begin{figure}[t]
\begin{center}
\includegraphics{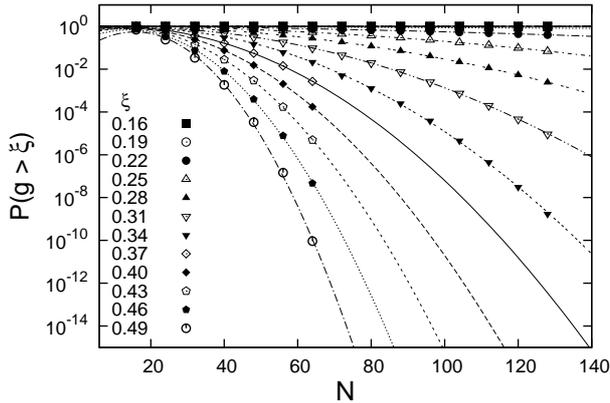}
\small
  \caption{Estimated $P(g>\xi)$s are shown as functions of $N$.
Each curve corresponds to different values of $\xi$. 
Data are well fitted by quadratic functions when $\xi \gtrsim 0.18$.}
  \label{fig:3}
\end{center}  
\end{figure}
\begin{figure}
\begin{center}
\includegraphics{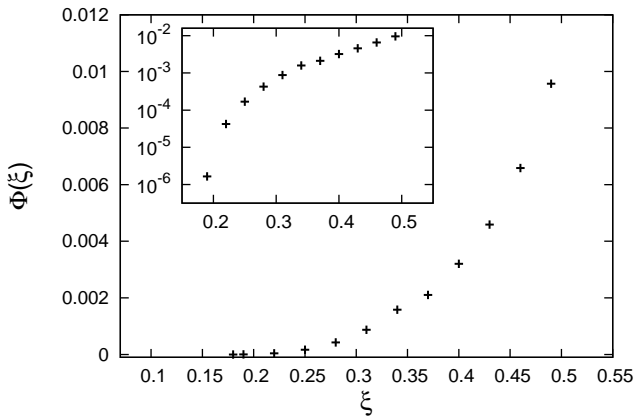}
\small
  \caption{The rate function $\Phi(\xi)$ is plotted.
The inset shows semi--log plot of $\Phi(\xi)$. 
In the region $\xi >0.18$, $\Phi(\xi)$ exponentially increases as $\xi$
 increases. $\xi=0.18$ is the point that $\Phi(\xi)$ becomes negative.}
  \label{fig:4}
\end{center}  
\end{figure}

To quantify decreasing rate of the tails of the distribution, we define the probability
$P(g > \xi)$ that $g$ is larger than $\xi$ by
\begin{equation}
 P(g > \xi)= \int _{\xi} ^{g ^{max}}\! p(g) \, dg.
\end{equation}
Here, we assume the probability $ P(g >g ^{max})$ is negligibly smaller
than $P(\xi \le g \le g^{max})$.
In Figure~\ref{fig:3},
estimated $P(g > \xi)$s are shown as functions of $N$.
For each $\xi$, $P(g > \xi)$ is well fitted
by a quadratic functions of $N$, when $\xi \gtrsim 0.18$.
This result indicates that $P(g > \xi)$ decreases for large $N$
as 
\begin{equation}
P(g > \xi) \sim \exp(-N^2 \Phi(\xi)),
\end{equation}
where the rate function $\Phi(\xi)$ is shown in Figure~\ref{fig:4}.
In the region $\xi \lesssim 0.16$, $P(g>\xi)$ is no longer a monotonic
decreasing function of $N$ and asymptotically approaches to unity for
large $N$.

\section{Concluding Remarks}
A method based on multicanonical Monte Carlo is introduced to the
estimation of large deviations in the spectral gap of random graphs.
By using this method, we calculate the distribution of the spectral gap
$g$ and the probability $P(g>\xi)$ for random 3-regular graphs.
While naive random sampling provides reasonable estimates of $P(g>\xi)$
only when $\xi$ is around the peak of the distribution $p(g)$,
the proposed method enables us to estimate $P(g>\xi)$ in a wide region of $\xi$
including extreme tails of the distribution.
We find that $P(g>\xi)$ behaves as
$P(g > \xi) \sim \exp(-N^2 \Phi(\xi))$ for large $N$, when $\xi \gtrsim 0.18$.
Our preliminary results indicate that a similar behavior
is also seen in the case of random $4$- and $5$-regular graphs,
suggesting that it is a general feature of random $k$-regular graphs.
The proposed method can be applied to calculations of large deviations in any
statics of any ensemble of random graphs.
In the case of non-regular graphs, the spectral gap are defined as the
smallest non-trivial eigenvalue of the Laplacian matrix
$L$~\cite{donetti2005entangled,donetti2006optimal};
hence we sample matrices $L$ instead of $A$ by using multicanonical
Monte Carlo.

Recent studies on Gaussian or Wishart random
matrices~\cite{dean2008extreme, vivo2007large}
showed that the probability of all eigenvalues being negative
decreases as
$\sim \exp(-N^2 \times \mbox{const})$ when the size $N$ of the matrices is large.
Our results are regarded as an extension of these results to the spectral
gap of random graphs.

\section{Acknowledgments}
We thank Prof.~M.~Kikuchi for his support
and encouragement.
This work is supported in part by Global COE Program (Core Research and
Engineering of Advanced Materials-Interdisciplinary Education Center for
Materials Science), MEXT, Japan.
All simulations were performed on a PC cluster at Cybermedia center, Osaka university.
\bibliographystyle{cpc}

\end{document}